 \documentclass[final,authoryear,5p,times,twocolumn]{elsarticle}
 
\usepackage{graphicx}

\usepackage{amssymb,amsmath}

\journal{Planetary and Space Science}

\begin{document}

\begin{frontmatter}



\title{Revisiting the lifetime estimate of large presolar grains in the interstellar
medium}


\author[label1]{Hiroyuki Hirashita}
\ead{hirashita@asiaa.sinica.edu.tw}
\author[label2]{Takaya Nozawa}
\author[label3]{Ryosuke S. Asano}
\author[label1]{Typhoon Lee}
\address[label1]{Institute of Astronomy and Astrophysics,
Academia Sinica, P.O. Box 23-141, Taipei 10617, Taiwan}
\address[label2]{National Astronomical Observatory of Japan, Mitaka, Tokyo 181-8588, Japan}
\address[label3]{Department of Particle and Astrophysical
Science, Nagoya University, Furo-cho, Chikusa-ku, Nagoya
464-8602, Japan}

\begin{abstract}
Some very large ($>0.1~\mu$m) presolar grains are sampled in
meteorites. We reconsider the lifetime of very large grains (VLGs)
in the interstellar medium focusing on interstellar shattering
caused by turbulence-induced large velocity dispersions.
This path has never been noted as a dominant mechanism of
destruction.
We show that, if interstellar shattering is the main mechanism
of destruction of VLGs, their lifetime is estimated to be
$\gtrsim 10^8$ yr; in particular, very large SiC grains can survive
for $\sim 1$ Gyr. The lifetimes obtained for VLGs are comparable
to the longest residence time derived for some presolar grains based on the
cosmic-ray exposure time. However, most presolar SiC grains show
residence times significantly shorter than 1 Gyr, which may
indicate that there is a more efficient mechanism than shattering
in destroying VLGs, or that VLGs have larger velocity dispersions
than 10 km s$^{-1}$.
We also argue that the enhanced lifetime
of SiC relative to graphite can be the reason why
we find SiC {among} $\mu$m-sized presolar
grains, while the abundance of SiC in the normal interstellar grains
is much {lower} than graphite.
\end{abstract}


\end{frontmatter}

% \linenumbers

\section{Introduction}
\label{sec:intro}

The elemental and isotopic compositions of presolar grains
found in primitive meteorites provide us with clues to
their origins. The peculiar isotopic ratios obtained for
presolar grains,
such as nanodiamonds, silicon carbide (SiC), and graphite,
indicate that they are formed in the environments around
particular types of stars {\citep{anders93,zinner14}}. SiC
is the species with the most detailed studies
because it is relatively easy to be separated from the
other substances in meteorites {and its high trace element
abundance allows us to perform isotope studies on a multitude of
elements. Also, SiC is one of the most abundant
$\mu$m-sized (in this paper, sizes are given as
radius)\footnote{{However, a factor 2 difference in size
does not matter for most of the statements in the Introduction.
Whenever precision is required, we use the word ``radius'' or
the notation $a$.}} species identified in presolar grains \citep{amari94}.}
Most of the SiC grains are defined as
mainstream grains, which have ${}^{12}\mathrm{C}/{}^{13}\mathrm{C}$
{isotope ratio = 10--100 (the solar ratio is 89) and enhanced
${}^{14}\mathrm{N}/{}^{15}\mathrm{N}$ ($>$272 = solar)}
\citep[e.g.,][]{amari01}.
The isotopic abundances in the mainstream SiC grains
indicate an asymptotic giant branch (AGB) star origin
{\citep[e.g.,][]{gallino94,lugaro03}}.
Presolar graphite grains are also identified: 
\citet{amari14} analyzed presolar graphite grains extracted from
the Murchison meteorite, and showed based on isotopic
ratios of various elements that some of them
{originated} from supernovae and others from AGB stars.
Theoretical studies of dust condensation in AGB star winds
also suggest that large ($\gtrsim 0.1~\mu$m) SiC grains form
\citep{yasuda12}. In this paper, we refer to grains larger than
$0.1~\mu$m as very large grains (VLGs).\footnote{The
term `large grain' or `big grain' is often used to
indicate the grain population which dominates the
far-infrared emission, and it refers to such a size range
as $\sim$0.01--0.25 $\mu$m \citep{desert90,li01}.
To avoid confusion with this convention, we refer to
large grains of interest in this paper as VLGs.}

After analysis of noble gas elements in SiC grains found in
primitive meteorites, \citet{lewis90} showed that
their isotopic and elemental ratios are consistent with
the values theoretically expected for
AGB stars. They also derived their presolar cosmic-ray
exposure age (or residence time) $\sim 135$ Myr based on the ${}^{21}$Ne
abundance. {\citet{heck09} inferred interstellar
residence times of presolar SiC as 3--1100 Myr for large
(diameter $\sim 5$--50 $\mu$m) presolar SiC grains.}
The residence times estimated may be uncertain
because recoil losses of Ne from $\mu$m-sized grains are
large. The measurement of spallation xenon (Xe), which is
much less affected by recoil loss, suggested shorter
cosmic-ray exposure ages $<$200 Myr \citep{ott05}.
\citet{gyngard09} estimated that the interstellar exposure
ages are 40 Myr--1 Gyr using $^6$Li excesses in presolar SiC grains.
These exposure ages give us clues to the lifetime (or
residence time) of VLGs in the interstellar medium (ISM).
Since AGB stars are one of the most important sources of
the interstellar dust \citep{ferrarotti06,ventura12},
the lifetime of dust grains formed in AGB stars is an
important quantity in considering the lifecycle of
dust in the ISM.

There have been some attempts of direct sampling of
interstellar dust using
dust-detecting instruments on spacecraft such as
\textit{Ulysses} \citep{grun94,baguhl95,kruger15},
\textit{Galileo} \citep{baguhl96},
\textit{Helios} \citep{altobelli06},
\textit{Cassini} \citep{altobelli03}, and
\textit{Stardust} \citep{westphal14}.
They indeed detected $\mu$m-sized grains
whose velocities point to their interstellar origin.
Interstellar grains
smaller than $\sim 0.1~\mu$m are
expelled by the solar radiation pressure and/or deflected by
the interplanetary magnetic field
\citep{levy76,sterken13}. Although the interpretation depends
on the modeling of the radiation pressure and gravity of
the Sun and the
magnetic field in the solar system \citep{sterken13},
the large abundance of VLGs measured by
the above spacecraft implies the existence of a
significant abundance of grains
larger than 0.1 $\mu$m in the ISM.

The lifetime of dust grains in the ISM is governed by destructive
processes, among which the most violent one is believed to be
supernova shocks. Assuming tight coupling between the motions
of gas and dust, \citet{jones96} argued
that shattering in supernova shocks is the major destruction path
for large $\gtrsim 0.1~\mu$m grains and estimated the typical
lifetime of the interstellar grains to be
$\sim \mbox{a few}\times10^8$~yr. However, VLGs are not necessarily
coupled with the gas motion in the ISM. Indeed,
\citet{slavin04} showed, solving explicitly the grain trajectories in
the presence of magnetic field, that decoupling of the
gas and dust is important
for grains with {radii} $>0.1~\mu$m in
supernova shocks. Therefore, VLGs may have lifetimes
much longer than $\sim 10^8$ yr, unless there is another
mechanism of limiting the lifetimes.

Although VLGs are decoupled from small-scale motions in the ISM,
they are coupled with a large-scale ($\sim$10--100 pc) turbulent and
magnetohydrodynamic motion in
the ISM \citep{yan04}. Based on an analytical kinetic theory of
dust grains, \citet{yan04} showed that grains larger than
$\sim 0.1~\mu$m can obtain random velocities as large
as 10 km s$^{-1}$ in the diffuse ISM. Using their results,
\citet{hirashita09} showed that VLGs are efficiently
disrupted by shattering in the diffuse ISM. Therefore,
even if VLGs escape from supernova shocks
by being decoupled from the gas motion, shattering in the
diffuse ISM is unavoidable.
This new destruction mechanism, not related to shocks, is
worth considering for the purpose of
constraining the lifetime of VLGs.

The existence of VLGs is also important in
the following two aspects. One is that the size range is
around the upper mass cut-off of grain radius of the
grain size distribution in the ISM.
A grain size distribution $n(a)\propto a^{-3.5}$
($a$ is the grain radius)
with an upper cut of grain radius at 0.25~$\mu$m
explains the Milky Way extinction curve
\citep*[][hereafter MRN]{mathis77}.
More elaborate models for the grain size distribution
also show a fall-off just above a similar grain radius
($\sim 0.25~\mu$m),
although it is not a rigid cut-off
\citep{weingartner01}. The other important aspect
of VLGs is that {a significant fraction of the total dust mass
is occupied
by grains whose sizes are around the upper cut (fall)-off
($\sim 0.1~\mu$m).
If we assume that MRN grain size
distribution in a grain radius range of 0.001--0.25 $\mu$m,
the VLG regime ($>0.1~\mu$m) contains
39 per cent of the dust mass.}
Therefore,
the lifetime of VLGs has a large impact on
the total dust abundance in the ISM.

In this paper, we reconsider the lifetime of VLGs, focusing on
shattering in the diffuse ISM. This process
has not been considered {to determine} the VLG
lifetime in previous studies. Moreover, as mentioned above,
VLGs may be decoupled
from supernova shocks. This means that the `classical'
picture in which the lifetime of dust grains is determined
by the shock destruction may not be valid for VLGs.
Therefore, the new destruction mechanism of VLGs
proposed in this paper will give a new upper limit for the
lifetime of VLGs. The lifetime estimated in this paper
is compared with the residence time of presolar grains
{inferred from isotopic measurements}.
Here,
the residence time indicates the time spent in the ISM
before the formation of the solar system.

The paper is organized as follows: we explain
the formulation adopted to estimate the lifetime of
VLGs in Section \ref{sec:model}. We show the
results in Section \ref{sec:result}. We discuss
the significance of the results in Section \ref{sec:discussion}.
Finally we conclude in Section \ref{sec:conclusion}.

\section{Models for shattering}\label{sec:model}

\subsection{Collision rate}\label{subsec:shat}

We consider a VLG once formed in stellar ejecta
and injected into the ISM.
We denote the radius and {root-mean-square (rms)} velocity
of a VLG as $a_\mathrm{VLG}$ and $v_\mathrm{VLG}$,
respectively.
{For simplicity, we assume that all VLGs have
the same velocity $v_\mathrm{VLG}$ since the velocity
distribution function of grains in turbulent medium is
unknown.}\footnote{{This assumption is not likely to affect the
results significantly. As shown in Section \ref{sec:result},
the destruction rate of a VLG is proportional to
$v^3$, where $v$ is the velocity of the VLG. If the
velocity distribution function, $f(v)$, follows the Gaussian weighted by the
phase-space volume ($f(v)=(l^3/\pi^{3/2})\exp (-l^2v^2)\cdot 4\pi v^2$,
where $l^2=3/(2\langle v^2\rangle )$, we obtain
$\langle v^3\rangle =1.23\langle v^2\rangle^{3/2}$, where
$\langle\cdot\rangle$ means the ensemble average. Thus, if
we represent the velocity $v_\mathrm{VLG}$ with the rms velocity
$\langle v^2\rangle^{1/2}$, $v_\mathrm{VLG}^3=\langle v^2\rangle^{3/2}$
approximates $\langle v^3\rangle$ well (unless we adopt a peculiar velocity
distribution function).}}
The interstellar turbulence is a viable
mechanism that causes the random velocities of grains
\citep[e.g.,][]{volk80}.
Since larger-sized grains are coupled with
larger-scale motions in the ISM and larger-scale turbulent
motions usually have larger velocity dispersions
\citep[e.g.,][]{yan04,ormel07},
we assume that VLGs obtain the largest
velocity dispersions, as large as
$\sim$10 km s$^{-1}$ (this value is taken from the
velocity dispersion of the diffuse neutral ISM in the
solar vicinity; \citealt{spitzer78,mathis00}).
This means that we can approximate the collisional
velocity between a VLG and a grain in the ISM
(referred to as an ISM grain in this paper)
with $v_\mathrm{VLG}\sim 10$ km s$^{-1}$, which is treated as a constant
parameter in this paper.

While a VLG travels in the ISM, it encounters
the ISM grains. The grain size distribution of the ISM
grains, $n(a)$, is defined so that $n(a)\,\mathrm{d}a$
is the number density of grains whose sizes are between
$a$ and $a+\mathrm{d}a$.
The rate at which
a VLG traveling in the ISM encounters the
ISM grains with radii between $a$ and $a+\mathrm{d}a$ is
denoted as $\mathrm{d}f_\mathrm{coll}$ and estimated as
\begin{eqnarray}
\mathrm{d}f_\mathrm{coll}=\pi (a_\mathrm{VLG}+a)^2v_\mathrm{VLG}n(a)\,\mathrm{d}a.
\end{eqnarray}

\subsection{Destroyed fraction}

In collision with an ISM grain,
a certain fraction of the VLG is
destroyed by shattering. In what follows, we adopt the formulation in
\citet{kobayashi10} to estimate the shattered fraction
 (the fraction of the volume
fragmented into smaller grains) of a VLG. They
estimate the shattered fraction of a grain as
\begin{eqnarray}
F_\mathrm{sh}=\frac{\phi}{1+\phi},
\end{eqnarray}
where $\phi$ is the impact energy normalized to the
threshold impact energy per mass for catastrophic
disruption,\footnote{Catastrophic disruption is defined as
a disruption event in which a significant fraction (half in the definition of
\citealt{kobayashi10}) of the grain is shattered.}
$Q_\mathrm{D}^\star$:
\begin{eqnarray}
\phi =\frac{v_\mathrm{VLG}^2}{2Q_\mathrm{D}^\star}\frac{y}{1+y},
\label{eq:phi}
\end{eqnarray}
with $y$ being the mass ratio, $m/m_\mathrm{VLG}$
($m$ and $m_\mathrm{VLG}$ are the masses of
the ISM grain and the VLG, respectively).

The contribution of the ISM grains with radii
between $a$ and $a+\mathrm{d}a$ to the shattered fraction of
{a} VLG per unit time is estimated as
$F_\mathrm{sh}\,\mathrm{d}f_\mathrm{coll}$. Therefore,
the total shattered fraction of the VLG increases
at rate $\mathrm{d}\Phi /\mathrm{d}t$ estimated as
\begin{eqnarray}
\frac{\mathrm{d}\Phi}{\mathrm{d}t} & = &
\int_{a_\mathrm{min}}^{a_\mathrm{max}}F_\mathrm{sh}
\frac{\mathrm{d}f_\mathrm{coll}}{\mathrm{d}a}\,\mathrm{d}a
\nonumber\\
& = &
\int_{a_\mathrm{min}}^{a_\mathrm{max}}\pi (a_\mathrm{VLG}+a)^2v_\mathrm{VLG}n(a)
\frac{\phi}{1+\phi}\,\mathrm{d}a.\label{eq:dPhidt}
\end{eqnarray}
Note that $\phi$ is a function of $a$.
If both the ISM grain and the VLG have the same
material density, $y=(a/a_\mathrm{VLG})^3$ in Eq.\ (\ref{eq:phi}).
We hereafter assume this expression for $y$.
The lifetime of the VLG is estimated as
\begin{eqnarray}
\tau_\mathrm{VLG}=1\left/\frac{\mathrm{d}\Phi}{\mathrm{d}t}\right. .
\label{eq:tau}
\end{eqnarray}

\subsection{Choice of parameter values}\label{subsec:parameter}

As mentioned in the Introduction, we are primarily interested in
SiC, which is well sampled for presolar grains, but we also
examine
silicate and carbonaceous (graphite) dust as they are known
to be representative
dust species in the ISM \citep[e.g.,][]{draine84}.
\citet{hirashita13} showed that $Q_\mathrm{D}^\star$ is estimated
using the critical pressure ($P_1$), above which the solid becomes
plastic, as
\begin{eqnarray}
Q_\mathrm{D}^\star\simeq\frac{P_1}{2s},
\end{eqnarray}
{where $s$ is the grain material density.}
We list $s$ and $P_1$ as well as the obtained $Q_\mathrm{D}^\star$
for each grain species in Table \ref{tab:param}.

Now we need to specify the grain size distribution of the
ISM grains, $n(a)$ (Eq.\ \ref{eq:dPhidt}).
We assume the MRN grain size
distribution $n(a)\propto a^{-3.5}$ with
the minimum and maximum grain radii
$a_\mathrm{min}=0.001~\mu$m and
$a_\mathrm{max}=0.25~\mu$m, respectively. This
grain size distribution is consistent with the
Milky Way extinction curve (MRN).
The results below do not change significantly
even if we adopt the grain size distributions proposed
for the Milky Way dust by \citet{weingartner01}.
The normalization
of the grain size distribution is determined by
\begin{eqnarray}
\mathcal{D}\mu m_\mathrm{H}n_\mathrm{H}=
\int_{a_\mathrm{min}}^{a_\mathrm{max}}\frac{4}{3}\pi a^3s
n(a)\,\mathrm{d}a,\label{eq:normalization}
\end{eqnarray}
where $\mathcal{D}$ is the dust-to-gas ratio,
$\mu =1.4$ is the gas mass per hydrogen nucleus,
$m_\mathrm{H}$ is the hydrogen atom mass, and
$n_\mathrm{H}$ is the number density of hydrogen nuclei.
To fix the number density of ISM grains, we always adopt
$s=s_\mathrm{ISM}=3$ g cm$^{-3}$ in Eq.\ (\ref{eq:normalization})
with $\mathcal{D}=0.01$ for the ISM grains.
We adopt $n_\mathrm{H}=0.25$ cm$^{-3}$ \citep{jones96}
for the typical density for the diffuse ISM,
keeping in mind that the grain lifetime is proportional
to $n_\mathrm{H}^{-1}$. We also assume that VLGs spend
most of their lifetime in the diffuse ISM before being
incorporated into
the molecular cloud from which the solar system was formed.

\begin{table*}
\centering
\begin{minipage}{160mm}
\caption{Dust properties adopted ($s$, $P_1$, and $Q_\mathrm{D}^\star$)
and calculated (the others).}
\label{tab:param}
\begin{center}
\begin{tabular}{@{}lcccccccc} \hline
Species & $s$ & $P_1$ & $Q_\mathrm{D}^\star$ &
$\eta$ & $\tau_\mathrm{VLG}\,^a$ & $\eta\tau_\mathrm{VLG}$
& $M_\mathrm{VLG}^\mathrm{AGB}$ & frac.$^b$ \\
 & (g cm$^{-3}$) & (erg\,cm$^{-3}$) & (erg\,g$^{-1}$)
 & & (yr) & (yr) & (M$_\odot$) & (per cent)
\\ \hline
Silicate & 3.3 & $3\times 10^{11}$ & $4.5\times 10^{10}$
  & $1.0\times 10^{-4}$ & $2.1\times 10^8$ & $2.1\times 10^4$
  & $6.3\times 10^4$ & 0.13\\
Graphite & 2.2 & $4\times 10^{10}$ & $9.1\times 10^9$
  & $2.2\times 10^{-4}$ & $4.4\times 10^7$ & $9.7\times 10^3$
  & $2.9\times 10^4$ & 0.058\\
SiC  &  3.1  & $1.7\times 10^{12}$ & $2.7\times 10^{11}$
  & $2.1\times 10^{-5}$ & $1.2\times 10^9$ & $2.5\times 10^4$
  & $7.5\times 10^4$ & 0.15\\
\hline
\end{tabular}
\end{center}
$^a$The lifetime is estimated at $a_\mathrm{VLG}=1~\mu$m.\\
$^b$Fraction to the total dust mass.
\end{minipage}
\end{table*}

\section{Result and analysis}\label{sec:result}

\begin{figure}
\includegraphics[width=0.45\textwidth]{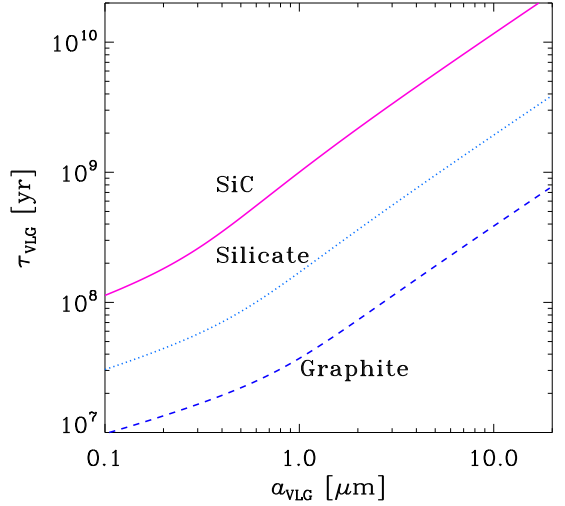}
\caption{Lifetime of VLG against shattering as a function
of radius $a_\mathrm{VLG}$. The solid, dotted, and dashed lines show
the results for SiC, silicate, and graphite, respectively.
\label{fig:lifetime}}
\end{figure}

We calculate the lifetime of VLGs, $\tau_\mathrm{VLG}$,
using the method described in the previous section.
In Fig.\ \ref{fig:lifetime}, we show $\tau_\mathrm{VLG}$
as a function of VLG radius $a_\mathrm{VLG}$ for the three dust
species. We observe that SiC {VLGs}
have lifetimes longer than
$10^8$ yr, {while silicate and graphite VLGs has significantly
shorter lifetimes than SiC VLGs.}
In particular, SiC grains with
$a_\mathrm{VLG}\gtrsim 1~\mu$m {(the radii appropriate for
actually sampled SiC VLGs)} survive for more than 1 Gyr
as long as shattering in the ISM is the main
mechanism of grain destruction.

The above result can be roughly reproduced by an
analytic argument as follows. As long as we consider VLGs,
$a_\mathrm{VLG}\gg a$
or $y\ll 1$ gives a good approximation for most
of the collisions. Therefore,
$\phi\simeq v_\mathrm{VLG}^2(a/a_\mathrm{VLG})^3/(2Q_\mathrm{D}^\star)$.
Since $a<0.25~\mu$m,
$\phi\lesssim 1$ holds for $a_\mathrm{VLG}\gtrsim 1~\mu$m, which means
that $\phi /(1+\phi )\simeq\phi$. With these approximations,
Eq.\ (\ref{eq:dPhidt}), combined with
Eq.\ (\ref{eq:normalization}), is reduced to the
following expression for
$\tau_\mathrm{VLG}=(\mathrm{d}\Phi /\mathrm{d}t)^{-1}$:
\begin{align}
\tau_\mathrm{VLG} &\simeq \frac{8a_\mathrm{VLG}Q_\mathrm{D}^\star s_\mathrm{ISM}}
{3v_\mathrm{VLG}^3\mathcal{D}\mu m_\mathrm{H}n_\mathrm{H}}\nonumber\\
&\simeq 1.2\times 10^9~\mathrm{yr}  %%1.17
\left(\frac{a_\mathrm{VLG}}{1~\mu\mathrm{m}}\right)
\left(\frac{Q_\mathrm{D}^\star}
{2.7\times 10^{11}~\mathrm{erg}~\mathrm{g}^{-1}}\right)
\left(\frac{s_\mathrm{ISM}}{3~\mathrm{g}~\mathrm{cm}^{-3}}\right)\nonumber\\
&\times
\left(\frac{v_\mathrm{VLG}}{10~\mathrm{km}~\mathrm{s}^{-1}}\right)^{-3}
\left(\frac{\mathcal{D}}{0.01}\right)^{-1}
\left(\frac{n_\mathrm{H}}{0.25~\mathrm{cm}^{-3}}\right)^{-1}.
\label{eq:tau_analytic}
\end{align}
This explains the values of $\tau_\mathrm{VLG}$ at
$a_\mathrm{VLG}\gtrsim 1~\mu$m in Fig.\ \ref{fig:lifetime}.
At $a_\mathrm{VLG}<1~\mu$m, the proportionality
($\tau_\mathrm{VLG}\propto a_\mathrm{VLG}$) does not hold
primarily because the approximation of $\phi\ll 1$ does not hold.
Yet, the above expression gives a good approximation within a
factor of $\sim$2 for the entire regime of VLG
(i.e., $a_\mathrm{VLG}>0.1~\mu$m).

The above results indicate that SiC VLGs
have long lifetimes against interstellar shattering
because of its high $Q_\mathrm{D}^\star$ value,
and that, as a grain becomes larger, its lifetime
becomes longer. This suggests that SiC VLGs tend to
preserve the memory of its formation site more than
silicate and graphite, which have shorter lifetimes.

\section{Discussion}\label{sec:discussion}

We discuss the implication of the lifetimes estimated
above. We focus on the comparison
with presolar
grains, whose residence times in the ISM can actually be
obtained. %%{with large analytical errors}.
We also discuss the expected abundance of VLGs
in the ISM.

\subsection{Comparison with presolar grains}\label{subsec:comparison}

The lifetime given above provides an upper limit for
the residence time of presolar grains in the ISM.
{Experimentally, the residence time of presolar grains
in the ISM can be constrained by using the excesses of
spallogenic isotopes, from which one can derive the fluence
of cosmic rays, hence the exposure time.}

\citet{gyngard09} estimated that the interstellar exposure ages
are 40 Myr--1 Gyr using $^6$Li {excesses in} presolar SiC grains
with radii $>1~\mu$m.
The range is consistent with the lifetime estimated above
(note that the lifetime gives an upper limit for the residence
time). Based on cosmogenic
${}^3$He and ${}^{21}$Ne,
\citet{heck09} derived the interstellar exposure ages
as $\sim$3--200 Myr for most of the sampled presolar SiC
grains with radii $>1~\mu$m, although a few grains
show {longer} interstellar residence times of 400 Myr to
1 Gyr. It seems that most of the sampled SiC
grains show much shorter residence times than the lifetime
estimated in this paper. Moreover, no trend of exposure ages
with grain size is observed in their sample, which is not
in {accordance} with our expectation that larger grains have
longer lifetimes. Since the {estimations} of exposure age
also {have uncertainties} in the
correction for recoil losses of the elements
{and
the poorly known Galactic cosmic ray flux at the formation
epoch of the solar system},
it is still premature to judge whether or not
we need to take these discrepancies seriously.

It is worth mentioning that the estimated VLG lifetime
is sensitive to the assumed velocity, $v_\mathrm{VLG}$
(Eq.\ \ref{eq:tau_analytic}). If we adopt a larger value
for the velocity, for example,
$v_\mathrm{VLG}=20$ km s$^{-1}$ as expected by the
acceleration by gyroresonance in the interstellar
magnetic field
\citep{yan04}, the lifetime of $\mu$m-sized SiC VLGs is
$\sim 10^8$ yr. Therefore, if an acceleration mechanism of
VLGs, such as gyroresonance, works in the ISM to
make the velocity dispersion of VLGs larger than that
of gas, the grain lifetime can be more consistent with
the residence times derived for a large part of the
above {samples. A} higher velocity
dispersion of VLGs than 20 km s$^{-1}$ is excluded
since the lifetime of $\mu$m-sized SiC grains becomes
shorter than 100 Myr, which contradicts the
residence times measured.

We further discuss other possibilities of explaining the short
exposure ages in Section \ref{subsec:processing}.

\subsection{Expected abundance of VLGs in the ISM}

We formulate the theoretically expected total mass
of VLGs, assuming that AGB stars are the
major production sources and that
interstellar shattering is
the major destruction mechanism. We consider the
total dust budget in the Milky Way.
The total mass of VLGs originating from AGB stars
in the Milky Way is denoted as
$M_{\mathrm{VLG},i}^\mathrm{AGB}$, where $i$ {stands for}
the grain species (silicate, carbonaceous dust, and
SiC). The time evolution of $M_{\mathrm{VLG},i}^\mathrm{AGB}$
is described as
\begin{eqnarray}
\frac{\mathrm{d}M_{\mathrm{VLG},i}^\mathrm{AGB}}{\mathrm{d}t}
=\left[\frac{\mathrm{d}M_{\mathrm{VLG},i}}{\mathrm{d}t}\right]_\mathrm{AGB}
-\frac{M_{{\mathrm{VLG},i}}^\mathrm{AGB}}{\tau_{\mathrm{VLG},i}},
\label{eq:dMVLG_dt}
\end{eqnarray}
where $[{\mathrm{d}M_{\mathrm{VLG},i}}/{\mathrm{d}t}]_\mathrm{AGB}$
is the production rate of VLGs of species $i$
in AGB star winds and $\tau_{\mathrm{VLG},i}$ is the
shattering time-scale of
VLGs in their collisions with the ISM grains.
We use the lifetime of VLGs estimated in Eq.\ (\ref{eq:tau})
for $\tau_{\mathrm{VLG},i}$.
The production rate of VLGs in AGB stars is
related to the star formation history, the statistics of
stellar mass (specified by the initial mass function),
and the total dust produced by individual stars
\citep[e.g.,][]{asano13}:
\begin{eqnarray}
\left[\frac{\mathrm{d}M_{\mathrm{VLG},i}}{\mathrm{d}t}\right]_\mathrm{AGB}
& = &
\int_{m_t}^{8~\mathrm{M}_{\odot}}\psi (t-\tau_m)m_i(m)\phi (m)\,\mathrm{d}m
\nonumber\\
& \simeq & \bar{\psi}\eta,\label{eq:AGBdust}
\end{eqnarray}
where $\psi (t-\tau_m)$ is the star formation rate (SFR) in the
Galaxy at $t-\tau_m$ ($t$ is the current age, and
$\tau_m$ is the lifetime of a star with mass $m$;
note that the dust production occurs at the end of stellar
life),
$m_t$ is the turn-off mass (the mass of a star whose lifetime
is equal to the age of the Galaxy; assumed to be 1 M$_\odot $ in
this paper),
$m_i(m)$ is the total mass of dust species $i$ produced by
the star with mass $m$, $\phi (m)$ is the stellar initial mass
function, and
$\eta\equiv \int_{m_t}^{8~\mathrm{M}_{\odot}}m_i(m)\phi (m)\,
\mathrm{d}m$. From the first to the second line in
Eq.\ (\ref{eq:AGBdust}), we approximate the SFR with the
time-averaged SFR ($\bar{\psi}$) assuming that the SFR does
not vary drastically as a function of time.
We do not consider a strong enhancement of SFR
at a particular time as discussed in
\citet{clayton03} (see also Section \ref{subsec:processing}).

For $m_i(m)$ we adopt the calculation of dust condensation in
AGB star winds in \citet{zhukovska08} for silicate,
carbonaceous dust, and SiC. Silicate in our calculation
includes all the species other than carbonaceous dust and
SiC, so that the total dust mass is equal to the
sum of the mass of SiC, graphite, and silicate.

Now we are only interested in
equilibrium states since the grain lifetime is significantly
shorter than the age of the
Galaxy. Even if this does not hold (i.e.,
$\tau_{\mathrm{VLG},i}\gtrsim t$), the equilibrium value gives
the maximum VLG mass expected in the Milky Way
since $\mathrm{d}M^\mathrm{AGB}_{\mathrm{VLG},i}/\mathrm{d}t>0$
holds (i.e., the VLG mass originating from AGB stars
still continues to increase until it reaches the equilibrium
value because the first term on the right-hand side in
Eq.\ \ref{eq:dMVLG_dt} dominates over
the second term).
Considering the equilibrium
$\mathrm{d}M_{\mathrm{VLG},i}^\mathrm{AGB}/\mathrm{d}t=0$
in Eq.\ (\ref{eq:dMVLG_dt}), combined with
Eq.\ (\ref{eq:AGBdust}), we obtain
\begin{eqnarray}
M_{\mathrm{VLG},i}^\mathrm{AGB}\simeq
\bar{\psi}\eta\tau_{\mathrm{VLG},i}.
\end{eqnarray}

In Table \ref{tab:param}, we show $\eta$,
$\tau_\mathrm{VLG}$ (estimated for $a_\mathrm{VLG}=1~\mu$m) and
$\eta\tau_\mathrm{VLG}$ for each dust species.
We observe that
{$\eta\tau_\mathrm{VLG}$ of SiC is as large as that of
silicate because the long lifetime of SiC VLGs compensates the
small yield of SiC in AGB stars.}
This explains the fact
that SiC is found in presolar grains {although
SiC is a minor dust species in the ISM compared with
silicate and graphite}; that is, if we sample
only $\mu$m-sized grains originating from AGB stars,
SiC VLGs are found {as easily as other species}
because of their long lifetime.

The SFR in the Galactic disk has been rather constant:
indeed, the current SFR is estimated as
$\sim 3$ M$_\odot $ yr$^{-1}$, and this level of SFR
explains the total stellar mass in the disk if
the disk age is $\simeq 10$ Gyr \citep{trimble00}.
Therefore, we assume that
$\psi\simeq\bar{\psi}\simeq 3$  M$_\odot $ yr$^{-1}$.
Multiplying this value with $\eta$ obtained above
(for $a_\mathrm{VLG}=1~\mu$m), we get
$M_{\mathrm{VLG},i}^\mathrm{AGB}$ for each species as listed
in Table \ref{tab:param}.

The total dust mass in the Milky Way is estimated to be
$\sim 5\times 10^7$ M$_\odot $
\citep{hirashita14}. We also list the ratio of
$M_\mathrm{VLG}^\mathrm{AGB}$ of each species to the total dust mass
in Table \ref{tab:param}. The fractions are
{on the} order of 0.1 per cent. Therefore, the existence of
VLGs in the ISM does not contradict the
observational properties reflecting the interstellar grains
such as extinction curves
and infrared emission because of the negligible
fractions.

However, these fractional abundances of VLGs cannot
be directly compared with the
fraction in primitive meteorites.
The actual fraction of sampled SiC grains in meteorites
{(up to 150 ppm; \citealt{zinner14})} is much smaller
than those values estimated above
probably because additional processing in the
solar system may have destroyed SiC grains by exposure to
hot ($\gtrsim$1,000~K) gas \citep{mendybaev02}.
{It is also interesting to compare the calculated and
measured silicate/SiC ratios.
We predict a silicate/SiC ratio of $0.13/0.15\sim 1$
(Table \ref{tab:param}) while
measurements of presolar grains indicate a ratio of $\sim 5$
\citep{leitner12}.
Although processing of SiC and silicate grains in the
solar system makes the direct comparison difficult,
it is encouraging that both theory and measurement show
silicate/SiC ratio of the same order of magnitude.}

\subsection{Grain processing mechanisms other than shattering}
\label{subsec:processing}

Sputtering in SN shocks %%with velocities $\gtrsim 100$ km s$^{-1}$
is the dominant mechanism of dust destruction for the ISM
grains \citep{dwek80,mckee89,jones94,nozawa06}.
However, \citet{jones11} pointed out that the destruction rate
of dust by supernova shocks is highly uncertain. Moreover,
as mentioned in the Introduction,
the motion of VLGs
tends to be decoupled from the shocks, which indicates
that some VLGs are likely to escape destruction by sputtering
in shocks. In this paper, we have considered
shattering in the diffuse ISM, not associated with supernova
shocks, and have given another constraint on the lifetime of VLGs.

VLGs may also form as a result of coagulation in the dense ISM
\citep{ormel09,hirashita14}.
The anomalous isotopic
compositions of presolar grains (Introduction) often
support the formation in specific types of stars.
{VLGs formed as a result of coagulation are expected to be compounds,
which do not show uniformity of composition.
Thus, it is
expected that VLGs of coagulation origin, even if they are
contained in a meteorite, are hard to
be identified as single presolar grains.
At the same time, VLGs of coagulation origin would
be less isotopically anomalous because coagulation of
many grains dilutes the peculiarity of
isotopic compositions.}
Although dust can also grow through the accretion of gas-phase
metals \citep[e.g.,][]{draine09},
VLGs cannot have been formed through accretion: as shown by \citet{hirashita11},
accretion only modifies a few nm of the dust surface, which is
too thin compared with the radii of VLGs.

\citet{bernatowicz03} studied 81 micrometer-sized {pristine}
presolar SiC grains. {They found that a large fraction of
the SiC grains are coated with an apparently amorphous, possibly organic
phase.} This may indicate that
the surface of SiC is affected by accretion in the ISM
or in the solar system. On the other hand, their SiC grain
surface studies
did not show strong evidence for cratering or sputtering.
They suggested that surface coatings protected the SiC
grains from destructive processing in the ISM.

As mentioned in Section \ref{subsec:comparison},
most of the sampled SiC presolar grains have lifetimes
much shorter than the theoretically estimated shattering
time-scale ($\sim$1 Gyr). \citet{ott05} suggested,
based on efficient amorphization in the ISM
inferred from {the high amorphous fraction (or low
crystalline fraction)} of silicate
by \citet{kemper04}, that a large part of interstellar
SiC grains are amorphized. Provided that amorphous SiC is easily
destroyed in the early solar system, the
amorphization time-scale is more relevant than the
shattering time-scale for the lifetime in the ISM
since the amorphized SiC grains are not sampled as
presolar grains.

Another solution for the short residence time of
presolar grains is to consider an enhancement of
AGB star dust production just before the solar
system was formed. This means that
an enhancement of SFR
(or a starburst)
occurred 1--2 Gyr before
the formation of the solar system \citep{ott05}.
In this star formation history, AGB stars originating
from stars with $m\sim 2$ M$_\odot$
form SiC dust grains just before the formation of the solar
system, so that their residence times are observed
to be short. \citet{clayton03} also explained the isotopic
ratios of the {mainstream} SiC presolar grains by
considering a starburst activated by a merger of
a metal-poor satellite galaxy with the Milky Way.

\section{Conclusion}\label{sec:conclusion}

We have investigated interstellar shattering as a mechanism
of determining the lifetime of very large grains
(VLGs with radii $a_\mathrm{VLG}>0.1~\mu$m) in the
interstellar medium (ISM).
Sputtering in supernova shocks, which {is}
considered to be the main destruction mechanism for
the interstellar dust, may not work for VLGs because
such large grains are decoupled from the gas motion on
the spatial scale of shocked regions \citep{slavin04}.
On the other hand, VLGs are coupled with a large-scale
turbulent motion in the ISM and attain random velocities
as high as $\sim 10$ km s$^{-1}$.
Therefore, we
have considered shattering of VLGs under successive
collisions with the grains present in the ISM (ISM grains).
Estimating the shattering time-scale for VLGs,
we have shown that SiC VLGs
have lifetimes as long as $\sim 1$ Gyr.

We have also compared the destruction time-scale obtained
above with the residence time of presolar SiC grains
sampled in meteorites. We referred to laboratory studies in the
literature for the residence time estimated from the cosmic ray
exposure time in the ISM using the {excesses of isotopes caused by
Galactic cosmic ray spallation}. Indeed, some presolar
SiC grains show residence times as long as 1 Gyr, which
is consistent with our estimate of lifetime above. However,
most presolar grains have residence times significantly
shorter than 1 Gyr, which may indicate that there is
a more efficient mechanism
than shattering in destroying VLGs, {that
a large abundance of dust was produced by AGB stars shortly before the
birth of the solar system}, or that VLGs have
larger velocity dispersions than assumed in this paper
(i.e., $>10$ km s$^{-1}$).

SiC VLGs have a longer shattering time than VLGs
composed of silicate and carbon.
The long lifetime of SiC VLGs may serve to enhance the
abundance of SiC in presolar $\mu$m-sized grains.

\section*{Acknowledgment}

We are grateful to S. Tachibana and the anonymous referees
for useful comments.
HH is supported by the Ministry of Science and Technology
(MoST) grant 102-2119-M-001-006-MY3.
TN is supported in part by the JSPS Grant-in-Aid for
Scientific Research (26400223).



%% References
%%

\end{document}